\documentclass[twocolumn, showpacs,prl]{revtex4}
\usepackage{graphicx}
\usepackage{dcolumn}	
\usepackage{bm}	

\newcommand\degrees[1]{\ensuremath{#1^\circ}}

\begin{document}

\title{Glass-like two-level systems in minimally disordered mixed crystals}

\author{J. P. Wrubel}
\author{B. E. Hubbard}
\author{N. I. Agladze}
\author{A. J. Sievers}
\affiliation{
Laboratory of Atomic and Solid State Physics\\
Cornell University, Ithaca, New York 14853-2501
}
\author{P. P. Fedorov}
\affiliation{Institute of General Physics, 117942 Moscow, Russia}
\author{D. I. Klimenchenko}
\author{A. I. Ryskin}
\affiliation{S. I. Vavilov State Optical Institute, 199034 St. Petersburg, 
Russia}
\author{J. A. Campbell}
\affiliation{Department of Physics, University of Canterbury, Christchurch, 
New Zealand}

\date{\today}

\begin{abstract}
THz spectroscopy is used to identify a broad distribution of two-level systems, characteristic of glasses, in the substitutional monatomic mixed crystal systems, Ba$_{1-x}$Ca$_x$F$_2$ and Pb$_{1-x}$Ca$_x$F$_2$. In these minimally disordered systems, two-level behavior begins at a specific CaF$_2$ concentration. The concentration dependence, successfully modeled using the statistics of the impurity distribution in the lattice, points to a collective dopant tunneling mechanism.
\end{abstract}


\maketitle

\newpage
The low temperature thermal properties of glasses and certain mixed crystals are dominated by a low lying distribution of excitations usually described within the phenomenological tunneling model as two-level systems (TLS) \cite{Pohl}. Although this TLS model is used to interpret the physics of disordered systems; justifications for either its assumptions or the well-known quantitative universality of fundamental physical characteristics have not been found \cite{Lubchenko,Enss}. A related soft-potential model may have a greater range of applicability \cite{Karpov,Gil}, but remains completely phenomenological. Other microscopic models which emphasize the potential energy landscape \cite{Wales} have not yet provided new insights into the low temperature anomalies. After the discovery of TLS in chemically disordered mixed crystals \cite{Walker, Kazanskii} it became apparent that progressively disordered crystals, in which, unlike glasses, the saturation of TLS behavior had not been reached, offered good opportunities for exploring the fundamental mechanisms responsible for the broad distribution of low energy states \cite{Pohl}. So far TLS behavior in crystals has been demonstrated only in systems for which disorder is produced either by a low symmetry component with orientational degrees of freedom, or by chemical disorder. Extensive work on fluorite mixed crystals shows broad TLS-like behavior for aliovalent dopants in the range of 1 - 45 mol \% \cite{Pohl,Fitzgerald1}. There the substitutional dopant cation introduces charge compensating interstitial anions while at the same time producing local strain and electric fields due to the different components. 

In this Letter we report on a THz study of the low temperature properties of two isovalent mixed crystal series where glass-like TLS have been observed. Both Ba$_{1-x}$Ca$_x$F$_2$ and Pb$_{1-x}$Ca$_x$F$_2$ have the fluorite structure and with some difficulty, single crystals of high quality can still be produced for $x\leq 0.1$. The appearance of TLS is consistent with the ideas that (1) the local dopant concentration varies statistically and (2) a critical concentration is required.

Due to the decomposition of both Ba$_{1-x}$Ca$_x$F$_2$ and Pb$_{1-x}$Ca$_x$F$_2$ solid solutions under cooling, a high cooling rate is used during crystal growth. Solid solutions Ba$_{1-x}$Ca$_x$F$_2$ ($x$ = 0.02, 0.04, 0.06, 0.08), crystal diameter 22 mm, crystal length 30 mm and Pb$_{1-x}$Ca$_x$F$_2$ ($x$ = 0.02, 0.04, 0.06, 0.08, 0.10), crystal diameter 10 mm, crystal length 30 mm were grown by the Stockbarger-Bridgeman technique. The melt was maintained for 1.5 hours at \textit{T} = 1400 \degrees{}C (Ba$_{1-x}$Ca$_x$F$_2$) and 2 hours at \textit{T} = 950 \degrees{}C (Pb$_{1-x}$Ca$_x$F$_2$). The crystals were grown at a rate of 10 mm per hour at \textit{T} = 1400 \degrees{}C (Ba$_{1-x}$Ca$_x$F$_2$) and \textit{T} = 950 \degrees{}C (Pb$_{1-x}$Ca$_x$F$_2$). After the crucible was transported over the gradient zone the furnace was switched off and the crystals were cooled to room temperature during 8 hours for Ba$_{1-x}$Ca$_x$F$_2$ and 4 hours for Pb$_{1-x}$Ca$_x$F$_2$. These rates are faster than the typical crystal growing cooling times of 16-24 hours. The resulting samples, which have large single crystal domains, are heavily strained, but optically transparent.

\begin{figure}[tb]
\includegraphics{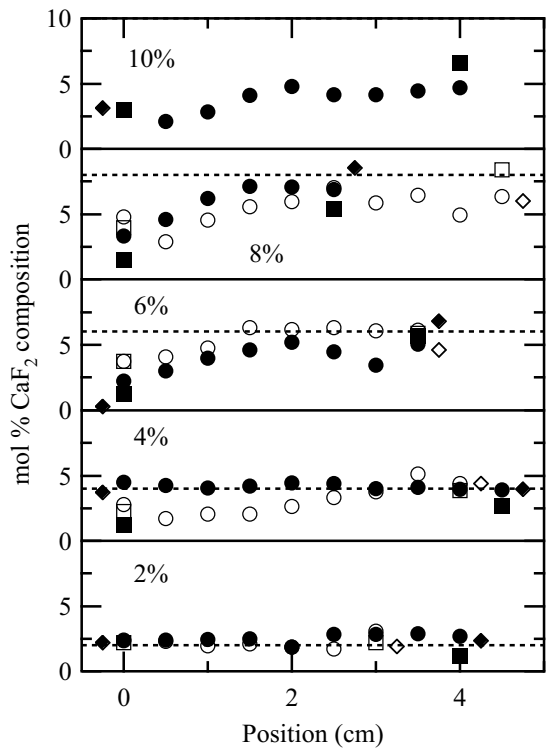}
\caption{\label{concentration}Raman and x-ray determination of the mole percent CaF$_2$ composition in the mixed crystals. Open points: Ba$_{1-x}$Ca$_x$F$_2$; filled points: Pb$_{1-x}$Ca$_x$F$_2$. Circles (squares) are from Raman T$_{2g}$ measurements along the sides (ends) of the samples. Diamond points are from powder x-ray scattering of end-cut pieces.}
\end{figure}

The dopant concentration was measured by powder x-ray diffraction of end chips as well as by Raman scattering along the entire length of each sample.  X-ray diffraction measures the average lattice constant, and Raman scattering measures the shift of the T$_{2g}$ symmetry phonon mode. Figure \ref{concentration} shows the Raman and x-ray data converted into the mol \% CaF$_2$ concentration, linearly extrapolated via Vegard's law \cite{Vegard}. The samples are identified in Table \ref{table:conc}. The Ba$_{1-x}$Ca$_x$F$_2$ samples showed significant variation over the sample length for greater than 2 \% CaF$_2$, consistent with the onset of the miscibility gap [10]; while the Pb$_{1-x}$Ca$_x$F$_2$ samples were inhomogeneous above about 4 \% CaF$_2$, again consistent with the phase diagram showing segregation starting at 4 - 6 mol \% CaF$_2$ \cite{Buchinskaya}. The mean CaF$_2$ concentrations were extracted by averaging the Raman data taken along the sides of each sample (Table \ref{table:conc}). Laue back-scattering measurements indicated that the samples are composed of few-mm single crystal domains with slight reorientations between neighboring crystals.

\begin{table}
\caption{\label{table:conc}Nominal and actual dopant concentrations $x_{Nom.}$ and $x$, in mole \% for individual mixed crystals; and the dielectric constant $\epsilon _0$ for each host.}
\begin{ruledtabular}
\begin{tabular}{cc D{+}{\,\pm\,}{3,3} |cc D{+}{\,\pm\,}{3,3} }
$\epsilon_0=6.94$& $x_{Nom.}$ & \multicolumn{1}{c}{$x$} &
$\epsilon_0=26.3$& $x_{Nom.}$ & \multicolumn{1}{c}{$x$}\\
\hline
BaCa1 & 2 & 2.2+0.5 & PbCa1 & 2 & 2.5+0.3 \\
BaCa2 & 4 & 3.1+1.1 & PbCa2 & 10 & 3.9+0.9 \\
BaCa3 & 6 & 5.4+1.1 & PbCa3 & 6 & 4.0+1.0 \\
BaCa4 & 8 & 5.4+1.2 & PbCa4 & 4 & 4.2+0.2 \\
              &    & \mbox{} & PbCa5 & 8 & 5.9+1.6
\end{tabular}
\end{ruledtabular}
\end{table}

\begin{figure}[tb]
\includegraphics{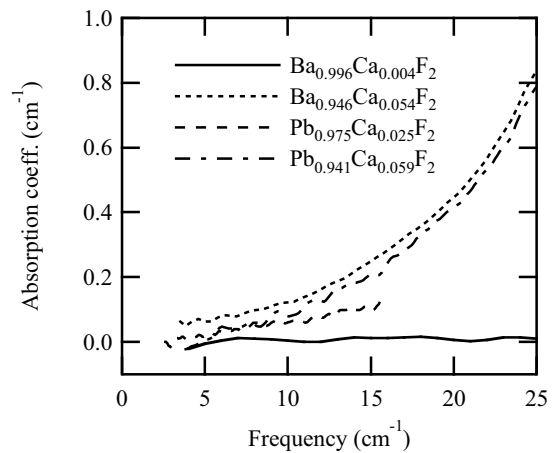}
\caption{\label{absorption}THz absorption coefficient of mixed crystals versus frequency at 4.2 K. No sharp defect-induced absorption peaks are observed. Sample concentrations are given in Table \ref{table:conc}.}
\end{figure}

Typical low temperature (1.5 K) THz absorption spectra over a decade in frequency are plotted in Fig. \ref{absorption} for some of these mixed crystal samples. These spectra show that no sharp impurity-induced absorption lines appear from 2 - 25 cm$^{-1}$, including the Ba$_{0.996}$Ca$_{0.004}$F$_2$ sample. Such sharp features are typical in the low concentration spectra of aliovalent mixed crystals \cite{Fitzgerald1,Fitzgerald2}.

When TLS make a significant contribution to the absorption, then increasing the sample temperature produces the characteristic bleaching observed in glasses \cite{Bosch}. The predicted temperature-induced change in the absorption coefficient $\alpha $, is \cite{Fitzgerald1}

\begin{equation}
\Delta \alpha (T_R,T)=-\frac{4\pi ^2 }{3c\sqrt{\epsilon _0 } }\left( \frac{\epsilon _0 +2}{3} \right) ^2 \omega \bar{P} (\omega)\,\,\mu _{b}^2 (\omega )\,\Delta p(T_R ,T), \label{alpha}
\end{equation}
where $\Delta p(T_R ,T)\equiv \left[ \tanh (\hbar \omega /2kT_R )- \tanh
(\hbar \omega /2kT) \right] $.  The quantity $\bar{P} (\omega )\mu _{b}^2 (\omega )$ is the optical TLS density-of-states (ODOS). Its frequency dependence is due to a cutoff in the high frequency tunneling energy and the onset of excited state transitions \cite{Fitzgerald1}. Taking the limit $\bar{P} (\omega \rightarrow 0)$ yields $\bar{P} (0)$, the constant spectral density of states per unit frequency range, $\mu _{b} $ is the electric dipole transition moment, \textit{c} is the speed of light, the term in parentheses is the local field correction, $\epsilon _0 $ is the dielectric constant, and $k$ is Boltzmann's constant.

\begin{figure}[tb]
\includegraphics{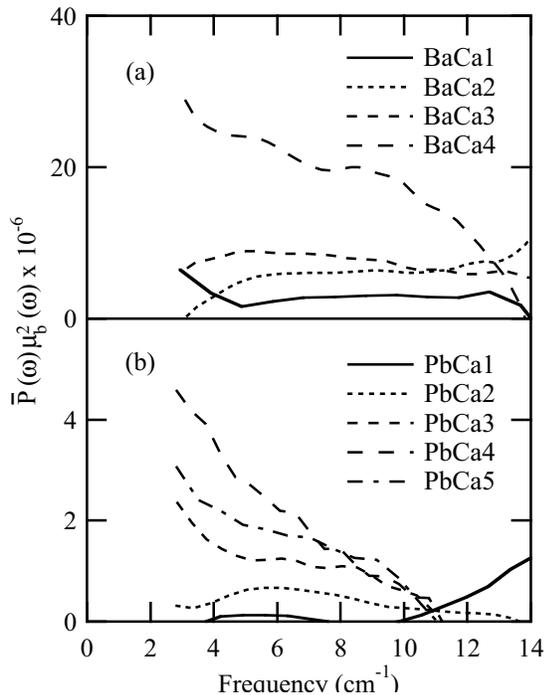}
\caption{\label{ODOSvsF}Measured TLS optical density of states of the mixed crystals versus frequency. The data are obtained from single parameter fits to the temperature dependence of the THz absorption coefficient. (a) Ba$_{1-x}$Ca$_x$F$_2$ and (b) Pb$_{1-x}$Ca$_x$F$_2$.}
\end{figure}

\begin{figure}[tb]
\includegraphics{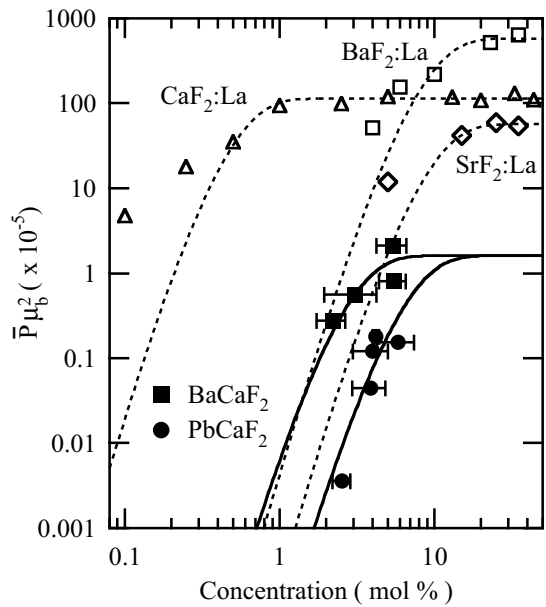}
\caption{\label{ODOSvsC}Critical concentration model fits to the TLS optical density of states for both aliovalent and isovalent mixed crystals. Data points are a weighted average from 3 - 10 cm$^{-1}$. The samples are identified in the figure.}
\end{figure}

The temperature dependent spectra between 1.5 and 10 K are used in conjunction with Eq. \ref{alpha} to extract the ODOS $\bar{P} (\omega )\mu _{b}^2 (\omega )$, which is graphed versus frequency for both series in Fig. \ref{ODOSvsF}. The Ba$_{1-x}$Ca$_x$F$_2$ spectra show two-level behavior increasing monotonically with concentration above a certain minimum value. For the lower concentration BaCa1 and BaCa2 samples, the TLS distribution is not yet completely glass-like since it does not extend below about 4 cm$^{-1}$.  Nonetheless, there is a significant range of two-level behavior and these data are used in Fig. \ref{ODOSvsC}. For the BaCa3 and BaCa4 samples, broad distributions of TLS are already present, although saturation has not been reached. The Pb$_{1-x}$Ca$_x$F$_2$ data shown in Fig. \ref{ODOSvsF}(b) demonstrate negligible ODOS for PbCa1 (2.5 \%) and a non-monotonic concentration dependence. For the Pb$_{1-x}$Ca$_x$F$_2$ samples, the onset of TLS behavior is very sharp and therefore sensitive to the precise sample concentrations. The maximum measured ODOS for Ba$_{1-x}$Ca$_x$F$_2$ is about one tenth of that in saturated mixed crystals and glasses \cite{Fitzgerald1}, and the ODOS in Pb$_{1-x}$Ca$_x$F$_2$ is about one hundredth of the saturated values.

Studies with doped alkali halide crystals have already demonstrated that for a given lattice containing a smaller size dopant cation that although single ions may have insufficient volume to tunnel, pairs of such ions can \cite{Greene}. In both of the systems studied here the dopant ion is the smaller; but there is no evidence of single ion or pair tunneling. We suggest that the TLS effect observed here is associated with a local statistical concentration of dopant ions exceeding a critical value $x_c$.

Since the probability that a given dopant ion is located in a specific subvolume is (subvolume/crystal volume), the characteristic subvolume for the disordered state needs to be identified. The first step is to introduce a Voronoi tessellation centered on all dopant sites that fills all space \cite{Okabe}. In this way two relevant volumes are defined; $V_i$ is the Voronoi polyhedron volume at site $i$, and $\bar{V} $ is the expectation value of $V_i$. The normalized Voronoi volume at site $i$ is then $z_i =V_i /\bar{V}  $. Assuming random dopant locations in the crystal and neglecting lattice discreteness, a Poisson Voronoi diagram follows. One feature of the Poisson diagram is that the probability density distribution for $z_i $ can be approximated by the gamma distribution \cite{Kumar}

\begin{equation}
p(z_i)=\frac{z_i^{a-1} }{b^{a} \Gamma \left( a\right) } e^{-z_i/b }, \label{poisson}
\end{equation}
where in 3-D $a=5.6333$ and $b=0.1782$.  Summing all Voronoi subvolumes less than a fixed critical value $V_c $, which is the fitting parameter, and defining $z_c =V_c /\bar{V}  $ gives the probability $W$ of locally exceeding a critical concentration

\begin{equation}\label{integral1}
W=\int\limits_0^{z_c}dz_i z_ip(z_i).
\end{equation}

In order to make contact with the ODOS, the saturation level $[ \bar{P} \mu _b^2] _{sat} $ is introduced as a second parameter so that the final expression is

\begin{equation}\label{integral2}
\bar{P} \mu _{b}^2 =\left[ \bar{P} \mu _{b}^2 \right] _{sat} \int\limits_0^{z_c}dz_i z_ip(z_i).
\end{equation}
The critical concentration can be computed from the fit parameter $V_{c} $ by $x_{c} =V_{host} /V_{c}  $, where $V_{host} $ is the volume of the host crystal primitive cell.

This critical concentration model is quite general and may be tested against all the THz data on TLS in mixed crystals. Fits have been carried out for three aliovalent impurity systems, namely CaF$_2$, BaF$_2$, and SrF$_2$ doped with LaF$_3$ and for the two isovalent systems studied here. The results are shown in Fig. \ref{ODOSvsC} and the parameters given in Table \ref{table:critical}. The model reproduced well both the delayed increase in TLS absorption with concentration as well as the saturation at higher concentration. The apparent underestimation by the model at low concentrations for CaF$_2$:LaF$_3$ may arise because narrow defect-induced tunneling states appear simultaneously with the incipient broad TLS absorption. In addition, clustering effects would result in higher than random appearance of regions with concentrations exceeding the critical value.

\begin{table}
\caption{\label{table:critical}Critical concentration model fit parameters.}
\begin{ruledtabular}
\begin{tabular}{c D{+}{\,\pm\,}{3,3} D{+}{\,\pm\,}{3,3}}
Crystals & \multicolumn{1}{c}{$[ \bar{P} \mu _b^2] _{sat}(\times10^{-5})$} & \multicolumn{1}{c}{$x_c$ (mol \%)}\\
\hline
(CaF$_2$)$_{1-x}$(LaF$_3$)$_x$ & 113+4 & 0.56+0.06 \\
(SrF$_2$)$_{1-x}$(LaF$_3$)$_x$ & 56+6 & 10+2 \\
(BaF$_2$)$_{1-x}$(LaF$_3$)$_x$ & 570+60 & 9.5+1.4 \\
Ba$_{1-x}$Ca$_x$F$_2$ & 1.6+0.3 & 3.2+0.3 \\
Pb$_{1-x}$Ca$_x$F$_2$ & \mbox{1.6} & 7.5+0.05 \\
\end{tabular}
\end{ruledtabular}
\end{table}

For the isovalent impurities studied here, the critical concentration model also shows qualitative agreement with the experimental data. The saturation level obtained by fitting the Ba$_{1-x}$Ca$_x$F$_2$ data is two orders of magnitude less than for the aliovalent mixed crystals. Most probably this is the result of a smaller dipole moment $\mu _b$ since the absorption by TLS created in isovalent systems is not enhanced by charge separation. The large uncertainty in the concentration of Pb$_{1-x}$Ca$_x$F$_2$ samples did not permit a reliable determination of $[ \bar{P} \mu _{b}^2] _{sat}$ so the Ba$_{1-x}$Ca$_x$F$_2$ value was applied.

It has been shown that two isovalent mixed crystals, Ba$_{1-x}$Ca$_x$F$_2$ and Pb$_{1-x}$Ca$_x$F$_2$ do contain significant although not saturated TLS optical density of states. The delayed but rapid onset of TLS behavior with increasing dopant concentration in these minimally disordered monatomic substitutional systems has been successfully modeled with a critical concentration parameter related to the statistics of the dopant distribution in the lattice. The other parameter, the saturation value, indicates that only a small fraction ($\sim 1/1000$) of the critical concentration regions contribute to the collective tunneling phenomenon. This fact suggests a complex energy landscape even for these simple lattice systems.

Finally, the observation of a step-like appearance of TLS with the number density, independent of the local concentration as soon as it exceeds the threshold value, implies that there are no intermediate states -- either TLS are present at the saturation level or they are absent. The success of this very simple model for mixed crystals points to a likely connection with the universality of TLS found in glasses.

\begin{acknowledgments}
JPW would like to thank the U.S. Department of Education for a graduate fellowship.  This work was supported by NSF-DMR Grant No. 0301035 and by the Department of Energy Grant No. DE-FG02-04ER46154.  PPF acknowledges support from the Russian Foundation of Basic Research (Grant No. 04-03-32836).
\end{acknowledgments}


\begin{thebibliography}{99}
\bibitem{Pohl}R. O. Pohl, X. Liu, and E. Thompson, Rev. Mod. Phys. \textbf{74}, 
991 (2002).
\bibitem{Lubchenko}V. Lubchenko and P. G. Wolynes, Phys. Rev. Lett. \textbf{87}, 195901 
(2001).
\bibitem{Enss}C. Enss, Physica B \textbf{316}, 12 (2002).
\bibitem{Karpov}V. G. Karpov, M. I. Klinger, and F. N. Ignat'ev, Solid State 
Commun. \textbf{44}, 333 (1982).
\bibitem{Gil}L. Gil, M. A. Ramos, A. Bringer, and U. Buchenau, Phys. Rev. 
Lett. \textbf{70}, 182 (1993).
\bibitem{Wales}D. J. Wales, Science \textbf{293}, 2067 (2001).
\bibitem{Walker}F. J. Walker and A. C. Anderson, Phys. Rev. B \textbf{29}, 5881 
(1984).
\bibitem{Kazanskii}S. A. Kazanskii, JETP Lett. \textbf{41}, 224 (1985).
\bibitem{Fitzgerald1}S. A. FitzGerald, A. J. Sievers, and J. A. Campbell, J. Phys.: 
Condens. Matter \textbf{13}, 2177 (2001).
\bibitem{Fedorov}P. P. Fedorov, I. I. Buchinskaya, N. A. Ivanovskaya, V. V. 
Konovalova, S. V. Lavrishchev, and B. P. Sobolev, Doklady Physical 
Chemistry \textbf{401}, 53 (2005).
\bibitem{Buchinskaya}I. I. Buchinskaya and P. P. Fedorov, Russian J. Inorg. Chem \textbf{43}, 
1106 (1998).
\bibitem{Vegard}L. Vegard, in \textit{Early Papers on X-ray Diffraction}, edited 
by Bijvoet, Burgers, and H\"{a}gg 1972), Vol. 2, 
\bibitem{Fitzgerald2}S. A. FitzGerald, A. J. Sievers, and J. A. Campbell, J. Phys.: 
Condens. Matter \textbf{13}, 2095 (2001).
\bibitem{Bosch}M. A. B\"{o}sch, Phys. Rev. Lett. \textbf{40}, 879 (1978).
\bibitem{Greene}L. H. Greene and A. J. Sievers, Solid State Commun. \textbf{44}, 
1235 (1982).
\bibitem{Okabe}A. Okabe, B. Boots, K. Sugihara, and S. N. Chiu, \textit{Spatial 
tessellations: concepts and applications of Voronoi diagrams} 
(John Wiley \& Sons Ltd, West Sussex, 1992).
\bibitem{Kumar}S. Kumar, S. K. Kurtz, J. R. Banavar, and M. G. Sharma, J. 
Stat. Phys. \textbf{67}, 523 (1992).
\end{thebibliography}
\end{document}